# Thermodynamic Behavior of a Model Covalent Material Described by the Environment-Dependent Interatomic Potential

P. Keblinski[1], M. Z. Bazant[2], R. K. Dash[1], and M. M. Treacy[3]

[1]*Materials Science and Engineering Department, Rensselaer Polytechnic Institute, Troy, NY 12180-3590*

[2] *Department of Mathematics, Massachusetts Institute of Technology, Cambridge, MA 02139*

[3]*NEC Research Institute, Inc., Princeton, NJ 08540*

## Abstract

Using molecular dynamics simulations we study the thermodynamic behavior of a single-component covalent material described by the recently proposed Environment-Dependent Interatomic Potential (EDIP). The parameterization of EDIP for silicon exhibits a range of unusual properties typically found in more complex materials, such as the existence of two structurally distinct disordered phases, a density increase upon melting of the low-temperature amorphous phase, and negative thermal expansion coefficients for both the crystal (at high temperatures) and the amorphous phase (at all temperatures). Structural differences between the two disordered phases also lead to a first-order transition between them, which suggests the existence of a second critical point, as is believed to exist for amorphous forms of frozen water. For EDIP-Si, however, the unusual behavior is associated not only with the open nature of tetrahedral bonding but also with a competition between four-fold (covalent) and five-fold (metallic) coordination. The unusual behavior of the model and its unique ability to simulate the liquid/amorphous transition on molecular-dynamics time scales make it a suitable prototype for fundamental studies of anomalous thermodynamics in disordered systems.

# I. INTRODUCTION

The idea of a first-order transition between two meta-stable disordered phases appears to be well established both conceptually and theoretically[1,2,3]. Moreover, such "liquid/liquid" (or "amorphous/amorphous") phase transitions have been observed experimentally for water[4], $Al_2O_3$-$Y_2O_3$ melts[5] and triphenylphosphite[6]. An analogous transition has also been postulated in the interpretation of the calorimetric data obtained in rapid heating of amorphous silicon and germanium thin films[7], although in such cases the direct observation of two coexisting disordered phases is hindered by rapid recrystallization.

The existence of a first-order amorphous/amorphous transition in the meta-stable extension of the phase diagram is believed to be responsible for unusual thermodynamic behavior, even in the stable region of the phase diagram, as evidenced by negative thermal expansion coefficients or a liquid of higher density than the crystalline solid. Due to its key importance in our life cycle, water is perhaps the most studied "unusual" substance. For example, liquid water has a maximum density at 4C and a minimum isothermal compressibility at 46C, which are both likely manifestations of thermodynamic singularities occurring in the supercooled region of the phase diagram.

Based on the results of molecular dynamics simulations it has been postulated that thermodynamic anomalies of water arise from the existence a second critical point, above which two metastable amorphous phases of ice become indistinguishable[8]. Recent molecular simulations of water[9] have revealed a comprehensive picture of various anomalies in the phase diagram. Specifically, the region in the phase diagram where density decreases upon cooling is embedded in a region where the diffusivity increases upon compression, while the broadest region of structural anomalies occurs where the structural entropy increases upon compression.

Anomalous structural, dynamical and thermodynamic behavior is typically exhibited by substances that form relatively open network structures associated with relatively specific chemistry, such as the hydrogen bond in water. In this context, especially from a theoretical perspective, silicon can be considered as a prototypical anomalous material. Despite being a pure elemental substance, silicon displays remarkably diverse coordination-dependent chemistry[10,11]. At low temperatures and pressures, strong covalent bonding characterizes the stable diamond crystalline phase (c-Si) and a meta-stable amorphous phase (a-Si), which both have open, four-fold coordinated structures. At high pressures, chemical bonding becomes more metallic, and various denser, over-coordinated crystal structures become stable, such as the six-fold coordinated -tin structure. At high temperatures, liquid silicon (l-Si) is also metallic, roughly six-fold coordinated and more dense than the crystal and amorphous phases. First principles electronic-structure calculations further reveal that the liquid has some remnant covalent bonding, with on average two out of six first neighbors forming transient covalent bonds[12,13,14]. These *ab initio* results are consistent with experiments[15] and molecular dynamics (MD) simulations[16] showing signs of partial covalent bonding in the structure factor of the supercooled liquid.

More than two decades ago it was predicted that the liquid/amorphous transition in silicon would be first-order[17,18], which has since has been confirmed in laser-heating experiments, albeit with extremely fast heating rates[19,20,21]. In experiments, however, it is very difficult to prevent crystallization of the supercooled liquid, so the liquid/amorphous transition has mainly been studied using MD simulations. The pioneering work of Car and Parinello first demonstrated that the liquid can be quenched directly into a realistic amorphous structure with 97% four-fold coordination in *ab initio* MD simulations with 64 atoms at constant volume[22,23]. Much less realistic amorphous structures (with over 25% coordination defects) have also been obtained by quenching small liquid samples using empirical tight-binding methods[24,25]. The practical limitations of short times and small system sizes, however, appear to preclude any detailed study of the liquid/amorphous transition with quantum mechanical methods, so instead empirical potentials have been used much more extensively.

A variety of artificial preparation methods, such as "indirect" quenching using modified potentials with enhanced angular forces[26,27], "bond-switching" randomization of the crystal[28], and the "activation-relaxation technique"[29], can produce reasonable amorphous structures (especially the latter two methods), but until recently only limited success has been achieved in direct simulations of liquid quench using empirical potentials. An early attempt to quench the liquid with the popular potential of Stillinger and Weber[30] (SW) failed to observe any transition to an amorphous phase, resulting only in a frozen liquid state[31]. With slower cooling rates, however, it was shown that the SW potential does in fact predict a reversible transition from the liquid to another, structurally distinct, low-temperature disordered phase[32,33,34]. This transition involves a nonzero latent heat and change in coordination, signifying a first-order transition, rather than a standard glass transition.

Although liquid quenching with the SW potential demonstrates the possibility of a first-order transition between two disordered phases of an elemental material, the low-temperature phase possesses over 20% coordination defects (more than one over-coordinated neighbor per atom). It thus bears little resemblance to real a-Si, whose structure is that of a random tetrahedral network of four-fold coordinated atoms with a very low concentration of coordination defects (typically <0.1%).

On a more fundamental level, the SW potential is not satisfactory as a model for phase transitions in silicon because it only has theoretical justification for one type of chemical bond, the ideal (or weakly distorted) $sp^3$ hybrid. As a result, the metallic bonding of the liquid is unrealistically modeled by severely distorted $sp^3$ hybrids having the same bond-length and bond-angle preferences as the crystal. Other covalent hybrids (e.g. $sp^2$, which is crucial for carbon) are not described at all. In other words, bonding preferences with the SW potential (and most others[35]) are completely insensitive to the local environment, which changes dramatically in phase transitions.

In an attempt to provide a more accurate description of bulk silicon and its phase transitions, the Environment-Dependent Interatomic Potential (EDIP) was recently proposed in Ref. 11 as a general model for covalent solids, taking into account the *ab initio* coordination dependence of chemical bonding obtained in Ref. 10 via exact inversions of cohesive energy curves. The parameterization of EDIP for silicon (EDIP-Si) provides a substantially improved description of the low-temperature structures of defects in the crystal phase, including point defects, stacking faults, and dislocations[36][37]. It also accurately describes the structure of a-Si and the dynamics of the liquid/amorphous[38] and crystal/amorphous[39] transitions, which makes it a natural choice for the present work.

In this paper we examine the thermodynamic properties of the EDIP-Si model. We demonstrate that, quite surprisingly, it exhibits a set of thermodynamic anomalies quite similar to those observed, or postulated to exist, for amorphous ice, while various aspects of high-temperature silicon are not well described. Nevertheless, we argue that the model is a useful general prototype for anomalous thermodynamic behavior, particularly in the context of amorphous/amorphous transitions.

The paper is organized as follows. In Section II the formulation of the EDIP for silicon is introduced, and existing results on its phase behavior are summarized. In Section III, an accurate thermodynamic melting point (crystal to liquid) for the model is reported, and in Section IV the thermal expansion of the crystalline, amorphous, and liquid phases is analyzed in detail. . Section V describes a study of the amorphous/liquid transition, and Section VI contains a discussion of the possible relevance of EDIP-Si for fundamental studies of various so-called "two-species models" and finally the conclusions are presented in Section VII.

## II. THE ENVIRONMENT-DEPENDENT INTERATOMIC POTENTIAL

In spite of many well-known limitations (reviewed in Refs. 11 and 35), the simple, pioneering potentials of SW and Tersoff[40] have traditionally been the most popular for MD simulations of silicon. In contrast, over thirty other empirical potentials, mostly involving large parameter sets or complicated functional forms with unclear physical interpretation, are much less popular. This experience suggests that the most successful approach to designing empirical potentials is to build the essential physics into a simple functional form containing only a few adjustable parameters, rather than brute-force numerical fitting. The latter method produces potentials with overly complex functional forms, which reduces computational efficiency and hinders physical interpretation.

In an effort to go beyond the practical limits of the guess-and-fit approach, recent work has provided *ab-initio* theoretical guidance to motivate a simple functional form for atomic interactions in covalent solids, the Environment-Dependent Interatomic Potential (EDIP) (Refs. 10, 11, and 36). Like most potentials for silicon, including SW, the EDIP functional form involves two-body radial forces to describe bond stretching and three-body angular forces to describe bond bending. Each of these forces depends strongly on an effective atomic coordination number, Z, which induces three-body and four-body interactions, respectively. The preferred radial distance lengthens and the bond strength decays with coordination number to describe the transition from covalent to metallic bonding. This is similar in spirit to the Tersoff potentials of Ref. 40, which vary the strength and length of a bond based on neighboring bond lengths and angles in a very different way, but in EDIP the local environment is defined from the perspective of the atom, rather than the bond.

A novel feature of EDIP is the strong environment-dependence of angular forces. The preferred bond angle shifts from the tetrahedral angle (109.471 degrees) at Z=4 to the hexagonal angle (120 degrees) at Z=3, to describe the transition from $sp^3$ to $sp^2$ covalent hybridization, respectively. As Z is increased above 4, angular forces are also substantially weakened to describe the transition to metallic bonding, which is well described by coordination-dependent radial forces with little angular dependence (e.g. embedded-atom potentials). Unlike the SW potential (and others), which can only describe tetrahedral semiconductors, EDIP can at least in principle describe more complex materials, as a result of its environment dependence. Indeed, a recent extension of EDIP to carbon (with an additional empirical description of -bonding beyond the *ab initio* justification for the original EDIP) appears to provide a good description of the different forms of amorphous carbon[41], which cannot be achieved with environment-independent potentials.

When fitted to various zero-temperature bulk phases and defect structures of silicon, EDIP-Si displays remarkable transferability for zero-temperature properties, including elastic constants, bulk crystal structures, point defects, stacking faults, and dislocation core reconstructions (Ref. 36). Unstable stacking-fault free energies at finite temperatures are also well described[42]. Also, quite remarkably, EDIP-Si is the first potential to predict a direct quench from the liquid into a high-quality amorphous structure. Even an extremely fast quench rate of -300 K/ps leads to a structure with 84% four-fold coordination, while at slower quench rates of -1 K/ps (still much faster than in laser-quenching experiments), a fairly realistic structure is produced with almost 95% four-fold coordination at room temperature. The vibrational properties of a-Si with voids[43] and crystal interfaces[44] have also recently been studied with EDIP-Si.

Occasionally, it is possible to quench the liquid with EDIP-Si to produce an "ideal" amorphous structure, completely free of coordination defects. This has recently been achieved with EDIP-Si using 64 atoms, a small system size which perhaps aids the exploration of phase space over MD time scales[45]. When relaxed with *ab initio* quantum-mechanical methods (in Ref. 45), the energy of the ideal structure goes from 0.15 eV/atom down to 0.10 eV/atom above the crystal ground state, which is remarkably low and (for the first time) well within the experimental energy range (< 0.19 eV/atom). The availability of an ideal amorphous structure generated by EDIP-Si makes it possible to isolate the effects of coordination defects on thermodynamic behavior below.

Consistent with its theoretical motivation, EDIP-Si also provides a useful tool to study phase transitions from the amorphous phase at moderately high temperatures in the range 800-1200K. For solid-phase epitaxial growth, the amorphous to crystal transition, EDIP-Si predicts an activation energy and a shape for the activated state in good agreement with experimental measurements of the interfacial velocity, and specific atomistic mechanisms for the transition can also be identified (Ref. 38). EDIP-Si has also been used to study ion-beam induced amorphization of the crystal and the subsequent material relaxation[46]. For amorphous melting, the amorphous to liquid transition, the interfacial velocity is also predicted in good agreement with experimental values, which is not the case with the SW potential (Ref. 37).

At still higher temperatures, however, EDIP-Si shows major differences with real silicon. In particular, the liquid at T = 1800 K has an average coordination of 4.5, well below the experimental value of 6.5. Moreover, the density is somewhat below that of crystal, when in reality the density increases by 10% upon melting. In spite of its lack of justification for metallic bonding in the liquid, these properties are well described by the SW potential. On the other hand, EDIP-Si curiously offers some qualitative improvement in the bond-angle distribution function (see Fig.1) by reproducing the auxiliary maximum at 60 degrees observed in *ab initio* simulations of the liquid (Refs. 13 and 14), although the primary maximum incorrectly occurs near the tetrahedral angle. In the original paper on EDIP-Si (Ref. 36), it was noted that the overly low density and coordination of the liquid may be artifacts of the short cutoff of the potential, which is appropriate for the low-temperature covalently bonded structures used in the fitting, but is perhaps too short for metallic phases like the liquid (which has also been said of the Tersoff potentials). This is perhaps also a reason or one of the reasons for the unusual behavior of EDIP-Si at high temperatures described below.

**III. CRYSTAL/LIQUID TRANSITION**

Despite the fact that many publications employing EDIP-Si potential have already appeared in the literature, no systematic studies of the crystal melting point have been reported, although it was mentioned in Ref 38 that its value is 1500 K. Part of the trouble in pinning down the melting point is the persistent misconception about how it should be determined. One problem is that, in a periodic defect-free crystal, nucleation of the liquid phase requires overheating well above the thermodynamic melting point. In the case of EDIP-Si, melting of the periodic crystal occurs around 2200 K, well above the experimental melting point of 1680 K. This difficulty can be removed by introducing either external interfaces[47] or grain boundaries[48] which eliminates the nucleation barrier. However, one needs to introduce planar interfaces, since otherwise the surface curvature shifts the apparent melting point. In the original attempt to establish the EDIP-Si melting point[49], a nano-cluster of 3,200 atoms was found to melt at 1370 ± 20 K. This is, however, a likely underestimation of the true bulk melting point, as evidenced by experimental results on small cluster melting temperatures[50].

To more accurately determine the bulk crystal melting point, we performed two independent sets of simulations on a system involving a planar liquid-solid interface. In both cases the simulation cell contains initially a perfect silicon crystal with dimensions along the *x*, *y*, and *z* coordinates equal to 5, 5, 20, respectively, in units of the silicon lattice parameter, $a_0$. In the first step, half of the system is melted by heating to 2500 K, while the rest of the atoms are kept at their crystalline positions. This results in an approximately $10a_0$ wide slab of liquid connected to a $10a_0$ slab of solid, with the normal to the crystal-liquid interface oriented along the (100) crystallographic direction. We note that, due to periodic boundary conditions, the system contains two liquid/solid interfaces in the simulation cell.

In the first set of simulations the system temperature is lowered to the vicinity of the likely melting point, and the whole system is allowed to evolve at constant energy without any constraints[51]. For the constant energy simulations throughout we use a time step of $8 \times 10^{-16}$ s to ensure good energy conservation over several hundred thousand steps (for constant temperature we use throughout a slightly larger time step of $1.4 \times 10^{-15}$ s). To avoid any effect of stress on the melting point we rescale and fix the lateral sizes (x and y) of the simulation cell according to the value of the lattice parameter of the crystal at given temperature. In the z direction we employ a variant of the standard constant-pressure algorithm to keep zz component of the stress tensor at zero value. This allows the accommodation of volume changes during melting or crystallization.

Plots of the coexisting liquid-crystal system temperature as a function of time are presented in Fig. 2. In one of the presented cases the temperature is initially above the melting point, which induces melting of the additional portion of the crystal. Due to the heat of fusion, however, this melting requires a lowering of the temperature as shown in Fig. 2. Finally, once the temperature reaches the melting point, the system is in equilibrium and no further change of temperature is observed, apart from small fluctuations. Fig. 2 also presents the temperature of a system that is initially placed below the melting point. In this case a portion of the liquid crystallizes, which injects additional thermal energy to the system and thus increases the temperature. Similarly, once the melting temperature is reached, only fluctuations around the true thermodynamic melting point are observed. As shown in Fig. 2, both constant-energy simulation runs yield, within the error bar, the same melting point of $1530 \pm 20$ K. This is, as expected, above the melting point of the 3,200-atom finite cluster and well below the stability temperature of the perfect crystal.

An alternative MD procedure relies on a series of constant temperature simulations of the liquid-crystal interface (Ref. 47). If the temperature is above the melting point, the interface moves towards the liquid with a temperature-dependent velocity, while below the melting point the crystal grows. Within the classical theory of nucleation and growth[52] it can be shown that the velocity of the crystal liquid interface, v, in the vicinity of the melting point, $T_m$, is proportional to

$$V \sim (T_m - T)\exp(-E/k_B T) \tag{1}$$

where E is an activation energy associated with the diffusion in the liquid phase (the atomic mobility in the crystal is essentially zero by comparison with the mobility of the liquid) and $k_B$ is the Boltzmann constant.

Figure 3 shows the value of the velocity of the liquid-crystal interfaces obtained from the analysis of the amount of liquid and crystal present in the simulation cell. This analysis relies on the monitoring the total system energy as a function of time. Due to the latent heat, when the crystal grows, the total energy decreases, while liquid growth is associated with an energy increase. Zero growth velocity indicates equilibrium. As shown in Fig. 3, the velocity versus temperature curve fitted to the formula given by Eq. (1) crosses the temperature axis at T=1520 ± 30K. This value of the melting point is in perfect agreement with that obtained from the constant energy simulations.

The fact that the melting point predicted by EDIP-Si is within 10% of the experimental value of 1685K is noteworthy because no properties of the liquid (or any other high-temperature structure) were included in the fitting database. Out of more than thirty potentials for silicon, no other predicts a reasonable melting point, with the exception of the SW potential and one other[53] which were explicitly constrained to reproduce the melting point during fitting. (The difficulty in predicting the melting point and other high-temperature material properties while maintaining realistic low-temperature properties is exemplified by the multiple parameterizations of the Tersoff potentials of Ref. 40.) In this light, it is also noteworthy that the experimental latent heat of melting (50.7 kJ/mol) is more accurately reproduced by EDIP-Si (37.8 kJ/mol) than by SW (31.4 kJ/mol) (Ref. 36). Nevertheless, the EDIP-Si liquid is quite different from real l-Si.

Returning to our MD simulations of the crystal-melt interface, in addition to the melting temperature, the fit to the velocity formula gives an activation energy, E = 0.65 ± 0.05 eV.

Interestingly, this energy for EDIP-Si is significantly higher than the corresponding value, 0.42 eV, for the SW potential (Ref. 47). As a result, the maximum growth velocity along the (100) crystallographic direction of ~5m/s with EDIP-Si is four times smaller than the value, 20m/s, obtained for the SW potential. Overall, the velocity versus temperature curve predicted by SW agrees well with the experimental results on (100) epitaxial growth of Si[54].

These results demonstrate that the SW potential provides a better description than EDIP-Si, not only of the liquid-phase structure, but also of the dynamics of crystal growth and melting. Since, unlike EDIP, the SW potential lacks any theoretical justification in the metallic liquid phase (discussed above), this may be related to the explicit fitting of the SW potential to the melting point. In any case, the difficulty of EDIP-Si in describing crystal-melt interfacial dynamics is perhaps due to the low coordination of the liquid, which may lead to a lower diffusivity and thus a higher activation energy and lower growth velocities. Since EDIP-Si provides a better description of the amorphous-liquid interfacial velocity than SW (see below and Ref. 38), however, it may also be that the problem in describing the crystal-liquid interfacial velocity is also due to unphysical properties of the EDIP-Si crystal at high temperatures, as we will discuss below.

## IV. THERMAL EXPANSION

### 1. The Crystal Phase

In addition to the melting temperature, thermal expansion represents one of the basic thermodynamic characteristics of a material. To obtain this property for the crystalline phase we equilibrate a 1,000 atom perfect crystal at a number of different temperatures and monitored the average specific volume over sufficiently long time periods (typically over 100,000 MD time steps at each temperature) to avoid excessive fluctuations. From such data we calculate the corresponding average values of the linear dimensions of the simulations box.

The size of the simulation box vs. temperature normalized to unity at T=0K is presented in Fig. 4 (open circles). It is striking to observe that, whereas the crystalline material initially expands with temperature, it reaches a maximum size at about 1200 K and then decreases. In other words, EDIP-Si predicts a *negative thermal expansion coefficient for the crystal at high temperatures*. This is contrary to the real Si crystal, which expands at high temperatures all the way to the melting point. Such a negative thermal expansion coefficient has been observed for ice[55] and a number of covalent materials including Si[56] at low temperatures, but at high temperatures only rather chemically complicated materials such as $ZrW_2O_8$[57] exhibit negative thermal expansion. Also, graphite has a slightly negative thermal expansion coefficient in the basal plane direction (a-axis), even at high temperature, but a much larger and positive coefficient in the normal to the basal plane direction (c-axis)[58].

To gain some insight into the origin of thermal contraction at high temperatures we monitor the temperature dependence of the distance between first, second, third and fourth nearest neighbor atoms. Such distances are obtained by monitoring time averaged radial distribution functions at various temperatures. Then we fit Gaussians and treat the mean positions as the average inter-neighbor distances.

The results of this analysis are also presented in Fig. 4. Whereas the first neighbor distance expands with temperature with increasing slope, the expansion of the second neighbor distance is smaller by roughly a factor of two. The third neighbor distance already shows contraction at high temperature, and the fourth neighbor distance follows the macroscopic behavior. We note that the fourth neighbors are at the distance corresponding to exactly one lattice parameter of cubic unit cell.

The preceding analysis indicates that *the unusual thermal contraction at high temperatures is not associated with simple bond contraction but rather with an attempt of the material to probe the "empty" interstitial space in its diamond crystal lattice*. In this way, the material can lower its free energy by increasing its entropy as it explores its rather open and low-coordinated crystal structure. The idea of an empty space be also understood by considering an eight-atom cubic unit cell of the diamond lattice: Four atoms reside at FCC lattice points while the other four atoms fill only half of the tetrahedral holes in the FCC lattice. This partial filling of the "holes" in the FCC lattice allows entropy-driven thermal contraction at high temperatures even when the bond length increases monotonically with temperature.

We also observe that the average bond angle between two vectors having a common origin on an atom and connecting two different nearest neighbors of the atom decreases monotonically with increasing temperature from the perfect tetrahedral angle of 109.4°. This observation is consistent with our interpretation of the thermal contraction observed for the model. However, we note that, for small perturbations, all forms of disorder on a regular tetrahedron will reduce the average bond angle, rendering this an insensitive measure to discriminate between different types of disorder.

## 2. The Amorphous Phase

The unusual property of negative thermal expansion is even more pronounced in the amorphous material. We first analyze a 1000-atom EDIP-Si amorphous sample prepared by a "slow" (-1K/ps) quench from the melt, which results in a relatively high quality random tetrahedral network with 96% four-fold coordinated atoms and 4% five-fold coordinated defects at zero temperature. We will refer to this model as "no-ideal" amorphous structure.

Figure 5 shows the thermal expansion data for this sample obtained in the same manner as described above for crystalline model. From the figure, it is apparent that *amorphous EDIP-Si contracts for all temperatures from absolute zero to theamorphous-liquid melting point*. At T=0 K, the coefficient of linear thermal expansion $(dL/dT)/L_c$ is -2.13 x $10^6$ 1/K, and the equilibrium length, L, at T = 0 K is 1.24% greater than that of the crystal. The first neighbor expansion of the amorphous material is very similar to that of the crystal, except at T = 0 K the average bond length is noticeably larger than the perfect crystal bond length. In contrast, the second neighbor distance is almost constant for the amorphous material and not monotonically expanding as in the case of the crystal. This indicates more pronounced bending of the angle between adjacent tetrahedral $sp^3$ bonds as they stretch, on comparison with the crystalline material.

The more pronounced thermal contraction of the amorphous material is perhaps associated with a shift in the competition between bond-length expansion and the lowering of the average bond angle caused by the structural disorder present in the amorphous phase (in contrast to only thermal disorder present in the crystal). Similar effects are observed experimentally for amorphous silica, which exhibits a lower thermal expansion than the crystalline phases. This explanation, however, requires some further refinement to account for defects.

The amorphous phase actually contains *two types of structural disorder*: (i) the random tetrahedral network of four-fold coordinated atoms, and (ii) a small concentration of five-fold coordinated defects. Since the preceding analysis does not distinguish between these two types of disorder, we also analyze the thermal expansion of the "ideal" 100% four-coordinated, periodic 64-atom EDIP-Si amorphous structure discussed is Sect. II. A detailed structural and electronic analysis is given in Ref. 45, but here we focus only on thermal expansion and compare with the results above for a slightly defective sample.

First, a 1728-atom ideal amorphous sample is created by replicating the periodic 64-atom sample 3x3x3 times and then randomizing the velocities. This defect-free sample is then gradually heated from T = 0.1K at 1 K/ps. Up to room temperature, the average number of defects remains less than one, so the structure may still be considered defect-free in this temperature range. Therefore, we fit the system length versus temperature for T < 273K to obtain the low temperature thermal expansion characteristics of the ideal amorphous structure. The zero-temperature volume is 5.4% larger than the crystal, which is greater than corresponding 3.8% for the "no-ideal" sample with 4% defects. Based on the above comparison we can assign an effective volume reduction of about one third of the atomic volume associated each five-fold coordinated defect. This is also consistent with a more comprehensive analysis of different defective samples in Ref. 45.

Above room temperature, the slowly heated ideal sample spontaneously forms a very small concentration of defects, reaching 0.1% at 500 K and 1% at 1000 K, as shown in Fig. 6. Since the presence of defects causes a local reduction in volume, it may seem that the gradual increase in defect concentration with temperature (in both the ideal and original samples) could explain the unusual negative thermal expansion reported above. It turns out, however, that this is only a secondary effect, at least at low temperatures. Instead, *negative thermal expansion is an inherent property of the ideal, defect-free amorphous structure*.

For T < 273 K, the constant-pressure linear thermal expansion coefficient of the ideal amorphous at T=0 is -2.86 x $10^{-6}$ 1/K, which is quite surprisingly 35% larger in magnitude than the same quantity for the no-ideal sample (with 4% defects). In other words, although the addition of new defects with increasing temperature lowers the overall volume, the presence of pre-existing defects at a given temperature actually inhibits thermal contraction. This is presumably due to the more compact local structure of over-coordinated defects compared to portions of the random tetrahedral network, which locally reduces the entropic driving force for contraction.

To further test this hypothesis, we measure a statistically averaged volume per defect at finite temperatures for ideal sample, using a technique from Ref. 45. In short we compare actual volume contraction in the temperature range of 300 K to 600 K with volume contractions extrapolated from low temperature regime (below 300K). Then by correlating the density of five-coordinated defects with additional volume contraction we obtain the excess volume per defect of - 8 +/- 3 $A^3$. This corresponds to a local volume reduction of about 1/3 in close agreement with the estimate obtained above for volume reduction associated with five-coordinated atoms at in zero temperature non-ideal amorphous structures. In conclusion, we now understand the negative thermal expansion of amorphous EDIP-Si as *a primary contraction of the random tetrahedral network plus a secondary contraction due to the creation of five-fold coordinated defects*.

### 3. The Supercooled Liquid Phase

In Fig. 7, we present the data for the specific volume of both crystalline and disordered phases over a larger temperature range normalized to the specific volume of the perfect crystal at T = 0 K. The results for crystalline material and low temperature amorphous material have already been discussed above. A dramatic decrease of the specific volume is observed above T > 1100 K. This temperature is associated with a transition of the amorphous solid into a denser and higher coordinated liquid, as we discuss in detail in the next section. The density of the liquid reaches a maximum at 1300 K and then decreases monotonically. Such a *density maximum in the supercooled liquid* is also predicted by the SW potential[59] and is associated with the fact that once the coordination number ceases to increase, increasing bond length leads to decreasing density. The liquid density maximum resembles the celebrated one of (stable) liquid water at 4 C, which has been attributed to the proximity to a second critical point in the metastable region of the phase diagram (e.g., see Ref. 3). However, this may not be the case here (see below).

It is interesting to note that 1300 K is close to the temperature where the crystal density reaches a minimum value. The higher density of the liquid over amorphous material is accompanied by increased coordination reaching the value of about 4.5 at T=1600 K, whereas it is just slightly above 4 in the amorphous phase. The coordination of the liquid is below the experimental value of 6.5, and this is perhaps the reason why at the l-c melting point of 1530 K the liquid is less dense than the crystal, whereas real l-Si is about 10% more dense.

## V. AMORPHOUS /LIQUID TRANSITION

### 1. Transition at Zero Pressure

Silicon is one of the simplest materials exhibiting two structurally different disordered phases, a covalent amorphous solid at low temperatures and a metallic liquid at high temperatures. From an empirical modeling point of view, the SW potential provides a good description of l-Si density (greater than that of the crystal) and coordination (around 6), albeit quite surprisingly since it has no theoretical justification for metallic bonding. The SW potential also allows direct MD simulations of transitions between the liquid and a low temperature amorphous structure, but the latter, have excessive number (20%) of coordination "defects". Furthermore, from a practical point of view the transition occurs very slowly on MD time scales with the SW potential.

On the other hand, EDIP-Si provides a superb description of a-Si but a rather poor description of the liquid density (less than the crystal) and coordination (4.5). In addition, the important advantage of EDIP-Si is that it transforms very easily between the liquid and amorphous phases on MD time scales. Therefore, we view EDIP-Si as a convenient model system to study first-order amorphous/amorphous phase transitions in a simple elemental material.

Following a standard approach to the study of transitions between the disordered phases, we first melt a 1,000 atom perfect crystal at high temperature and rapidly quench it to 1,300 K at zero pressure, where the liquid is still stable. We then quench the sample at a much slower rate of 1 K/ps (which still quite fast on experimental time scales) until the transformation to the amorphous structure is complete. Finally, we heat the sample back to the starting temperature of the cooling cycle, once again going through the melting transition.

The energy of the system as a function of temperature in the heat cycle just described is shown in Fig. 8. At the high and low temperature ends, the heating and cooling curves overlap indicating the presence of equilibrium liquid and amorphous phases, respectively. In the vicinity of the transition, i.e., in the 1000 - 1200 K temperature range, the liquid persists upon cooling, while the amorphous phase persists up to higher temperatures upon heating. Both heating and cooling curves have an s-shape, a signature of the underlying first-order transition (as previously observed for the SW potential in Refs. 32-34). The underlying reason for the double s-shaped heat cycle curve is the jump in energy associated with the latent heat of amorphous material melting. Similar behavior is observed for the density and average coordination.

To establish the amorphous melting temperature more precisely, we perform constant-energy, constant pressure, simulations analogous to those used to calculate of the crystal melting point above in Section III. A coexisting liquid-amorphous system is created by first preparing the amorphous and liquid components in separate simulations at 900 and 1200 K respectively. Then, the liquid and amorphous parts are joined by sintering under slightly positive pressure, and equilibrated at T=1,150 K for 20,000 MD steps. One such system consists of 10,000 atoms and has dimensions $5a_0$ x $5a_0$ x $\sim 50a_0$. Following this preparation procedure, the periodic system containing coexisting amorphous and liquid phases, and two flat interfaces between them, is subjected to constant-pressure and constant-energy simulations for several hundred thousand MD steps during which we monitor average values of a number of system properties.

By monitoring the slice-by-slice time-averaged density, energy and coordination across the liquid/amorphous interface, we observe a liquid with higher coordination, density, and energy coexisting with the amorphous material. From the energy profiles, we establish that the latent heat of the liquid/amorphous transition is not zero, signifying a *first-order phase transition*. The precise value of the latent heat is $0.10 \pm 0.01$ eV/atom. This is unlike in the standard glass transition which is believed to be either a second-order phase transition or a sudden change in kinetic behavior, involving zero latent heat in either case.

For EDIP-Si, as in the case for real Si, the amorphous/liquid transition also involves a discontinuous change in structure, as evidenced by a sudden change in coordination (from 4.07 to 4.22). The bond-angle distribution also makes a sudden change from a narrow, single peak at the tetrahedral angle for the amorphous phase to the broad, bi-model distribution of the liquid shown in Fig. 1. These discontinuous changes in local symmetry are typical for a first-order transition. The unusual feature of the amorphous/liquid transition, however, is that it does not involve any change in macroscopic symmetry, since both phases have no long-range order.

**2. Pressure-Temperature Phase Diagram**

The equilibrium melting temperatures resulting from constant-energy simulations of coexisting liquid and amorphous phases for several different pressures are shown in Fig. 9. As expected, due to the higher liquid density, increasing the pressure promotes this phase and thus lowers the transition temperature. At zero pressure, we estimate the amorphous/liquid transition temperature to be $1170 \pm 20$ K, which is 23% below the crystal melting temperature of 1520 K. Experimentally[60], the amorphous melting point for silicon is estimated at 1420 K, which is roughly 16% below the crystal melting point of 1685 K. Our result is close to 1200 K, a value obtained in recent constant temperature simulations of amorphous/liquid interface of the EDIP-Si using velocity versus temperature method (Ref. 38). We note also that below negative 0.5 GPa the system becomes unstable and cavities are formed, making it practically impossible to explore further into the negative-pressure region of the phase diagram.

The presence of the two structurally distinct disordered phases suggests the existence of the second critical point in the meta-stable region of the phase diagram where amorphous and liquid phases become indistinguishable. In an attempt to determine the location of this possible critical point we monitor the average coordination in the coexisting liquid and amorphous phases at various pressures. The selected results presented in Fig. 10, however, show no signs of criticality. The average coordination of both liquid and amorphous material increases with pressure, but the difference between them is almost constant and equal to about 0.15 at all studied pressures. This indicates that either our simulations are still far away from the critical point (which may be even located in inaccessible negative pressure region) or are too short to reach equilibrium, considering the slow dynamics near the critical point.

Future simulations, with significantly longer simulation times and system sizes, may perhaps resolve the issue of second critical point for EDIP-Si. If one were found for such a model elemental substance, it would be quite remarkable because a second critical point has previously only been suggested by MD simulations of water. Such simulations, however, are beyond the scope of this work and would require significantly more computational power.

### 3. The Role of Defects

The atomic structure of defects in real amorphous silicon has been a subject of considerable controversy in the literature [45,61,62,63]. Experimental studies invariably suggest that three-fold coordinated coordinated atoms ("dangling bonds") are the primary defects in the random tetrahedral network (e.g. see Ref. 62 and references therein). As pointed out by Pantelides, [61] some experimental observations are also consistent with over-coordinated defects, such as the five-fold coordinated "floating bond". Indeed, all theoretical calculations of bulk amorphous silicon, using both *ab initio* and empirical methods, invariably predict over-coordinated defects (e.g. see Ref. 45 and references therein). The reason for this glaring discrepancy between theory and experiment is still unclear, but it may be related to the presence of voids with reconstructed surfaces. In any case, this controversy is not relevant in the present context because we are studying EDIP-Si only as a convenient model system and do not intend to draw conclusions about real silicon. For the purposes of this paper, therefore, we always refer to five-fold coordinated atoms as the "defects" in the amorphous phase.

In order to understand the role of amorphous defects (five-fold coordinated atoms) in the phase transition, we have also slowly heated the ideal, defect-free 1728-atom sample described above (obtained by liquid quench) until it melts. As the sample is heated above room temperature, it develops a small concentration of five-fold coordinated defects, whose concentration is roughly given by Z-4, where Z is the average EDIP coordination of the sample, as shown in Fig. 6. Even at high temperatures in the range 1000-1200K, the sample has considerably fewer defects (1-3%) than the non-ideal sample (4% at T = 0K increasing to 7% at the l-a transition temperature) and resists transformation into the liquid. Eventually, however, the sample undergoes a rather dramatic first-order phase transition, as indicated by sudden changes in coordination, volume, and energy at T = 1360 K, as shown in Fig. 11.

Presumably as a result of perfect four-fold coordination, *the ideal amorphous sample significantly overheats before melting, just like the ideal crystal*. This is quite unusual for a disordered structure, which one would normally expect to contain plenty of nucleation sites for the liquid. Instead, the random tetrahedral network of the amorphous phase has remarkable stability, comparable to that of the ideal crystal, when the defect concentration is reduced.

An intriguing alternate possibility suggested by our simulations is that the melting point and other thermodynamic properties cannot be unambiguously defined for "the amorphous phase", because it is by definition meta-stable (i.e. out of equilibrium), and hence does not correspond to a unique set of allowable thermodynamic states. Indeed, many properties of real a-Si, such as the average coordination, are known to depend sensitively on sample history in experiment. Certainly at low temperatures (roughly below 800 K) the amorphous structure is effectively frozen out of equilibrium and can only relax very slowly without completely exploring its phase space.

At higher temperatures (roughly above 1000 K), however, defects are much more mobile, and one would expect that a quasi-equilibrium is attained. In this regime, the defect concentration is likely to be able to reach a well-defined steady value for a given temperature and pressure, and hence it should be meaningful to discuss thermodynamic averages and a unique melting point. This interpretation is supported by the two independent measurements of the EDIP-Si a-l melting point described above (ours and that of Ref. 38) which yield the same value.

In our simulation of heating the ideal amorphous structure, the unusually low defect concentration at high temperatures may simply be the result of an overly short MD simulation time, which does not allow the system to reach its equilibrium defect concentration. Indeed, a closer look at the simulation data shows a sharp change in slope for the curves of coordination, volume, and energy versus temperature near the equilibrium melting point of 1170 K determined above. Above this temperature, it may be that ideal sample was heated too quickly to allow for a proper equilibration of the defects, which may be necessary before nucleation of the liquid and subsequent melting can occur. Note that the defect concentration reaches 5 % when the superheated sample actually begins to melt, which is close to the 7 % equilibrium concentration in the simulations using "non-ideal" amorphous material.

## VI. TWO STATE MODEL

The fact that the amorphous phase of the EDIP Si is predominantly four-fold coordinated and high-temperature liquid phase is a roughly equal mixture of four- and five-coordinated atoms suggests that EDIP-Si may be used in fundamental studies representing what is sometimes called the "two species model" [64], the "two-state model" [65], or the "two-level model"[66]. In such models, one assumes the existence of two "species" in one component phase, A and B, which can be interconverted via a physical or chemical reaction. The equilibrium concentrations of species A and B, which in the EDIP-Si model corresponds to four-fold and five-fold coordinated atoms, respectively, depend on temperature and pressure, consistent with our analysis of the ideal and defective amorphous samples during melting simulations. If the two species are assumed to be in non-ideal solutions, the system may exhibit a first order phase transition[67], as is the case for the system studied in this work.

To graphically illustrate the concept of the two species model for EDIP-Si, in Fig. 12 we present atomic positions of four-fold (open circles) and five-fold (solid circles) coordinated atoms in a 40Å x 40Å x 5Å thick slab of liquid at T=1400 K and zero pressure cut from a larger structure. At this temperature the composition is about 1/3 five coordinated and 2/3 four coordinated atoms. It also appears that the five fold coordinated atoms tend to cluster, which is consistent with the non-ideal nature of the A-B solution. A more quantitative analysis indicates that the average cluster size is in the range of 1-2 nm at this temperature and pressure.

This clustering is very similar to that observed in molecular simulations of the SPC/E model[68] of glassy water, where low-density and high-density amorphous water tends to form similar clusters at the microscopic level[69]. Another striking resemblance of the EDIP-Si liquid to amorphous water is in the bond-angle distribution function shown in Fig. 1, which exhibits two maxima at 60 degrees and 100 degrees, associated with five- and four-fold coordinated atoms, respectively. The orientational order parameter in SPC/E glassy water also exhibits a very similar bimodal distribution[70]. The analogy between the unusual thermodynamic behavior of water and silicon has been extensively discussed by Angell and his collaborators (Ref. 59. The results of our studies underscore similarities between water and silicon despite apparent differences in chemical complexity. Like silicon, water has a locally tetrahedral arrangement, with adjacent molecules connected via the two lone pairs, and the two hydrogen atoms. Perhaps, this locally tetrahedral character and associate open network structure of water and silicon are principle factors in their anomalous thermodynamic properties.

Perhaps the most interesting aspect of the amorphous/liquid transition of the model studied in this paper is associated with the relatively fast transition dynamics allowing for direct MD dynamical studies of this process. To illustrate this point, in Fig. 13 we present the atomic positions and the coordination profiles of a system that was initially prepared as a liquid and then slowly cooled and equilibrated at the l-a transition temperature using EDIP-Si. Instead of homogeneously transforming the liquid into the amorphous structure, the sample undergoes a *complete spontaneous phase separation into liquid and amorphous regions*. This indisputable evidence for a nonzero interfacial tension provides additional proof of the first-order nature of the phase transition and opens possibility of future studies of dynamical behavior of this two state system.

## VII. CONCLUSIONS

In this paper we have examined a wide range of thermodynamic properties of the EDIP model for silicon including the crystal, liquid and amorphous phases. We have demonstrated that, compared to other empirical potentials, EDIP-Si provides an excellent description of the low-temperature crystal and amorphous phases. We have also accurately calculated the crystal and amorphous melting points with EDIP-Si using two different methods and have found that they are predicted remarkably well (contrary to previous published results). The crystal melting point is notoriously difficult to predict with empirical potentials, and, in contrast, previously it has only been reproduced by explicit fitting. On the other hand, we have also shown that EDIP-Si does not properly describe other thermodynamic properties and the structure of the high-temperature phases. In particular, both the crystalline and amorphous phases exhibit negative thermal contraction at high temperatures, and the liquid has a considerably lower coordination than real liquid silicon.

Nevertheless, we have shown that EDIP-Si provides a simple and computationally efficient model system to study anomalous thermodynamic behavior, including a first-order transition between two structurally distinct disordered phases. (For comparison, simulations of water require the much more demanding computation of long-range Coulomb interactions.) We expect that future simulations of the dynamics of this phase separation, and the associated coarsening process, may provide important insights into the dynamics of such unusual phase transitions. Finally, our results show that EDIP-Si has some high-temperature thermodynamic properties, which are not only qualitatively, but also quantitatively, similar to those of water, thus suggesting a general link between tetrahedral bonding and anomalous thermodynamic behavior.

## IV. ACKNOWLEDGMENT

We thank Prof. Cornelius Moynihan, Prof. H. Eugene Stanley, and Prof. Michael J. Aziz for an invaluable discussion on the amorphous/amorphous transition and the two-state model. This work was supported by the by National Science Foundation under Grant No. DMR 00-74273

.

**FIGURE CAPTIONS**

Figure 1.

Bond angle distribution functions of the liquid at T=1800 K for the EDIP-Si potential (solid line), Stillinger-Weber potential (dotted line), and *ab-initio* model from Ref 13 (dashed line). In each case, the cutoff is taken to be the first minimum of the pair correlation function, g(r), beyond its first maximum, which for EDIP-Si is around 2.85 A. The peaks at 60 and 100 degrees in the EDIP-Si bond-angle distribution are associated with five-fold and four-fold coordinated atoms, respectively. If a smaller cutoff is used (below roughly 2.45 A) the peak at 60 degrees is reduced, and the main peak shifts closer to the tetrahedral angle, indicating the covalent character of shorter bonds.

Figure 2.

Temperature vs. time dependence of the coexisting crystal and liquid obtained in the constant energy simulations. Regardless of the initial temperature, the system will reach the melting temperature in equilibrium (indicated by the dotted line).

Figure 3.

The velocity of a crystal-liquid interface as a function of temperature. Positive and negative velocity corresponds to crystal growth and melting respectively. The solid line is the fit of the data to the formula given by Eq. 1. The horizontal and vertical dotted lines indicate the melting point.

Figure 4.

Linear thermal expansion of crystalline EDIP-Si. Also shown is the temperature dependence of the average distances between first, second, third and fourth nearest neighbors.

Figure 5.

The same as Fig. 3, but for an amorphous model material obtained by quench from the liquid.

Figure 6.

The concentration of five-fold coordinated defects (given by Z-4, where Z is the EDIP-Si coordination number) versus temperature in the random tetrahedral network of four-fold coordinated atoms of our "ideal" amorphous sample.

Figure 7.

Specific volume of the crystalline, liquid and amorphous phases as a function of temperature normalized to unity for the zero temperature perfect crystal.

Figure 8.

The energy of the disordered system in the vicinity of the amorphous/liquid transition for cooling and heating. The double s-shaped curve indicates an underlying first-order transition.

Figure 9.

Amorphous/liquid transition temperature as a function of pressure obtained from a constant energy simulation of coexisting phases with a planar interface between them.

Figure 10.

Average atomic coordination in the coexisting liquid and amorphous phases as function of pressure.

Figure 11.

Liquid-amorphous transition for an "ideal" amorphous structure. (a) EDIP coordination number, Z (b) specific volume (per atom) and (c) energy.

Figure 12.

Positions of four-fold coordinated (open circles) and five-fold coordinated (solid circles) atoms in a 40Å x 40Å wide, 5Å thick slab of supercooled liquid at 1400 K.

Figure 13

Top panel: Positions of four-fold coordinated (open circles) and five-fold (plus a few six-fold) coordinated (solid circles) atoms in a 1024-atom slab of EDIP-Si material in equilibrium at the liquid/amorphous melting point, which has undergone a spontaneous phase separation into an amorphous layer and a liquid layer parallel to the left and right free surfaces. (The periodic sample is replicated once in the vertical direction for clarity.) Bottom panel: Spatial distributions of four-fold and five-fold (plus some six-fold) coordinated atoms in the direction perpendicular to the two layers for the structure in the top panel.

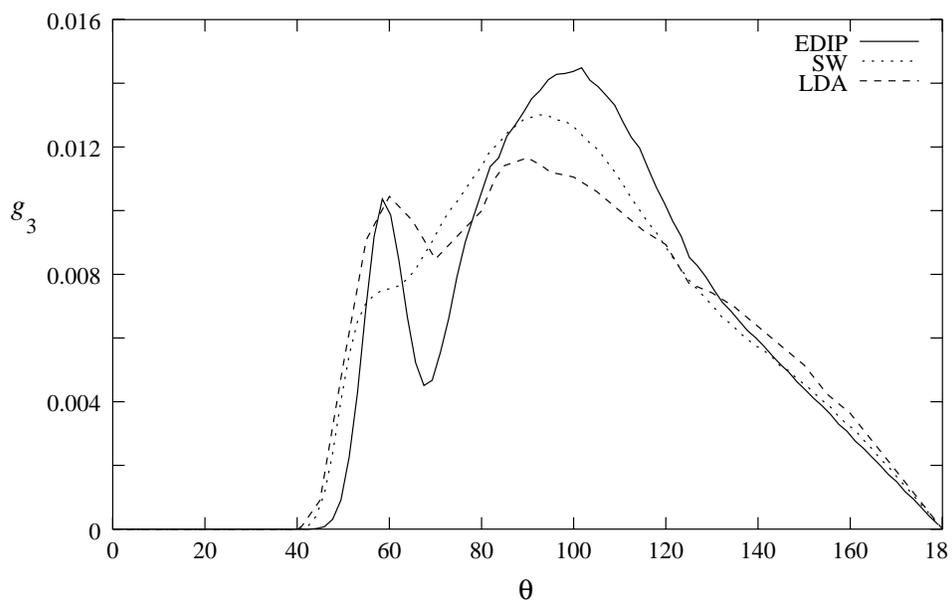

Figure 1. Keblinski *at al.*

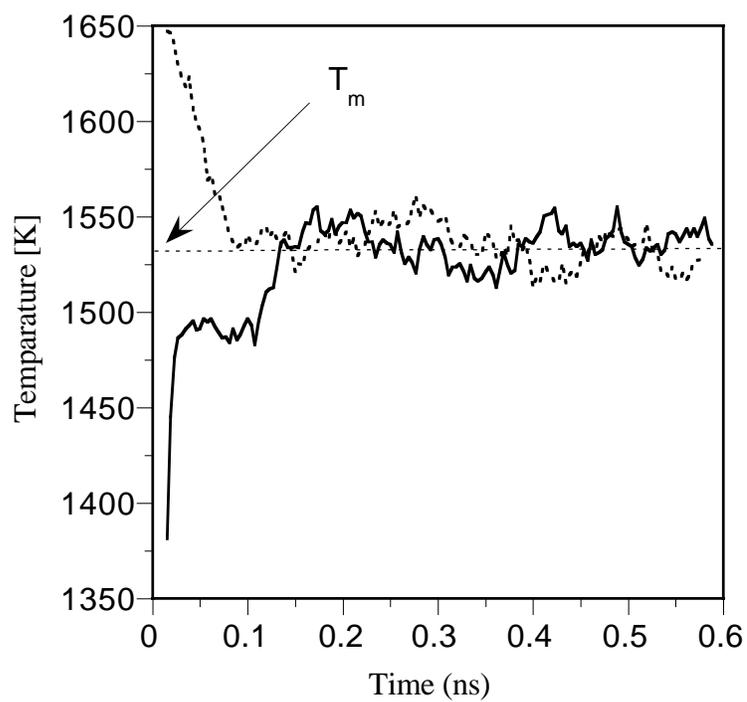

Figure 2. Keblinski *at al.*

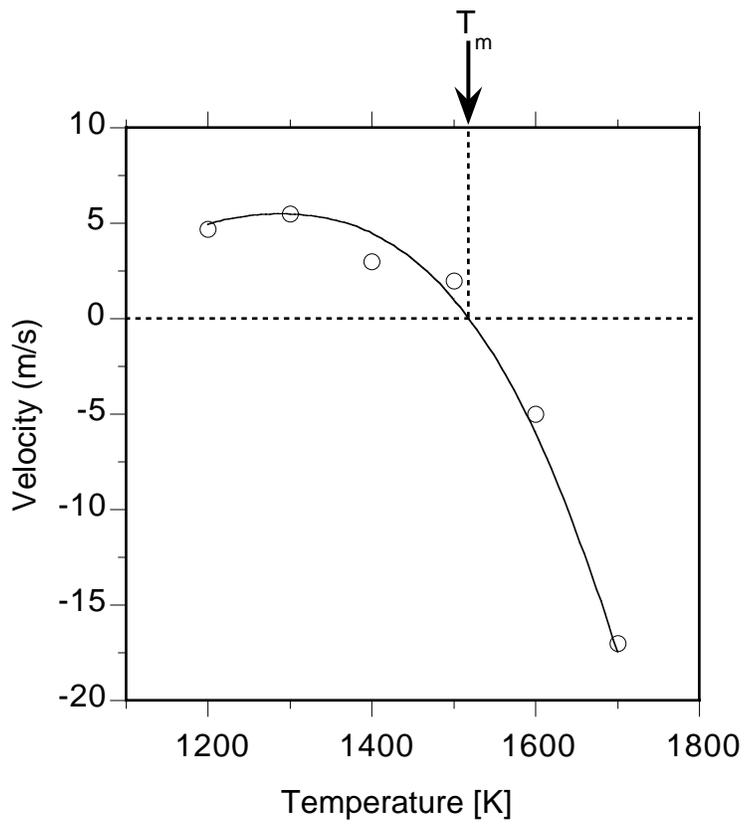

Figure 3.                                                                 Keblinski *at al.*

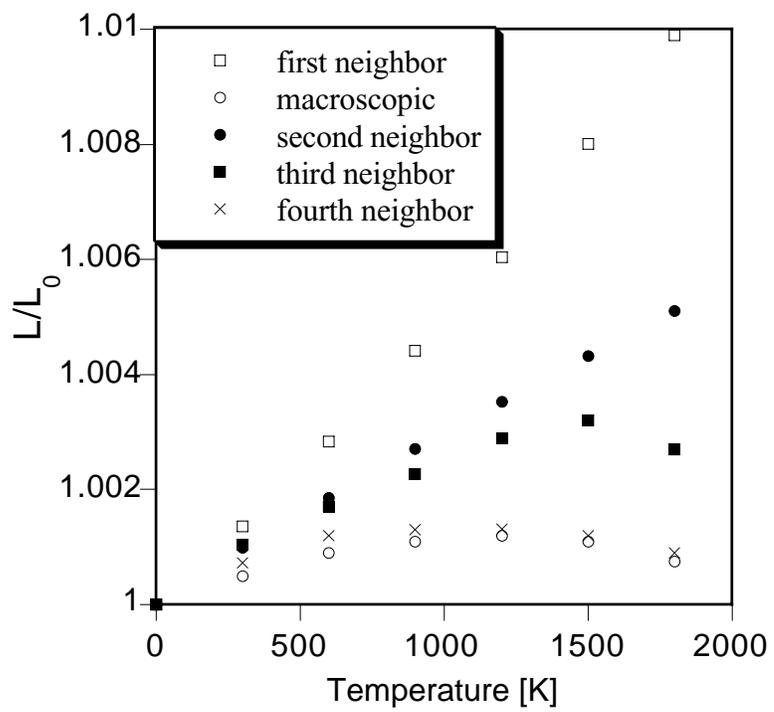

Figure 4. Keblinski *at al.*

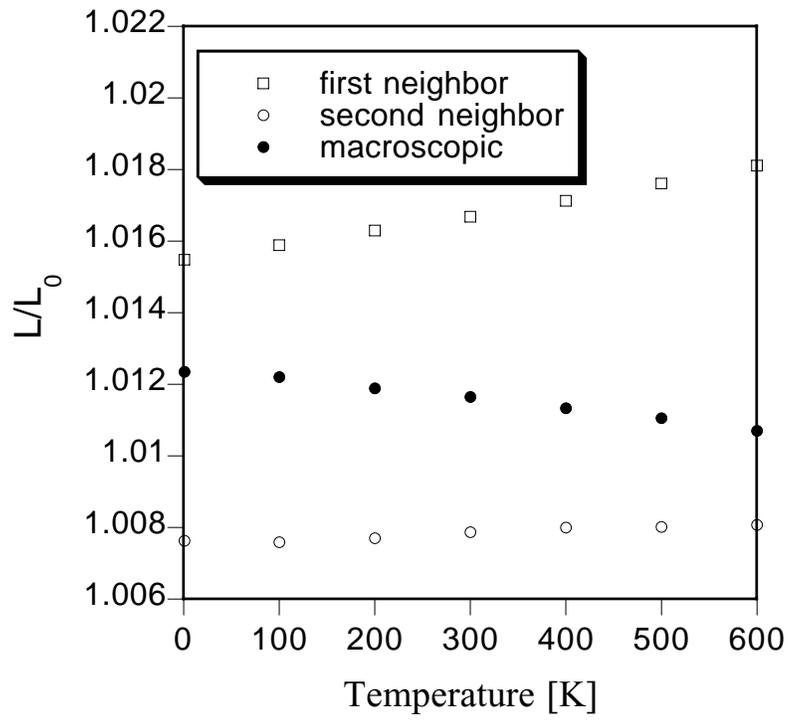

Figure 5. Keblinski *at al.*

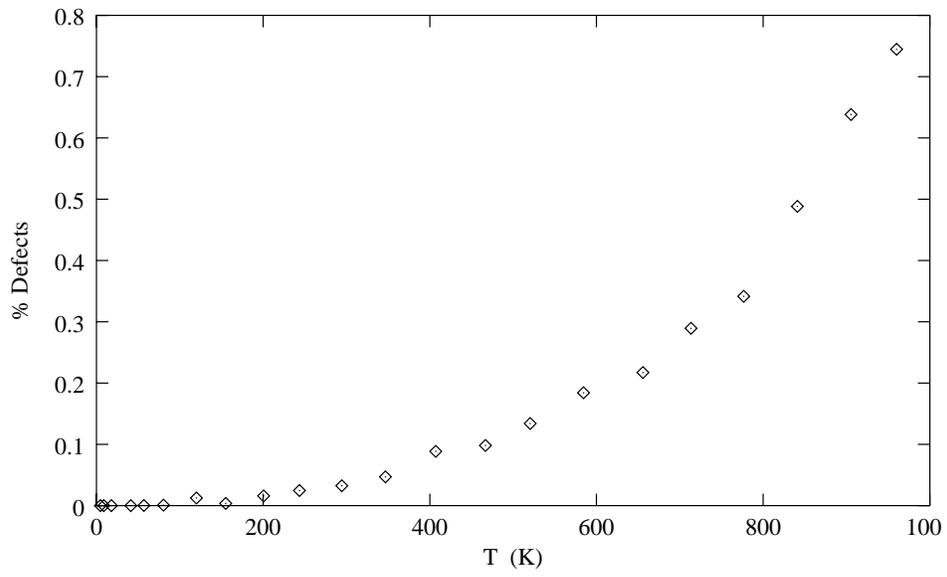

Figure 6.                                             Keblinski *at al.*

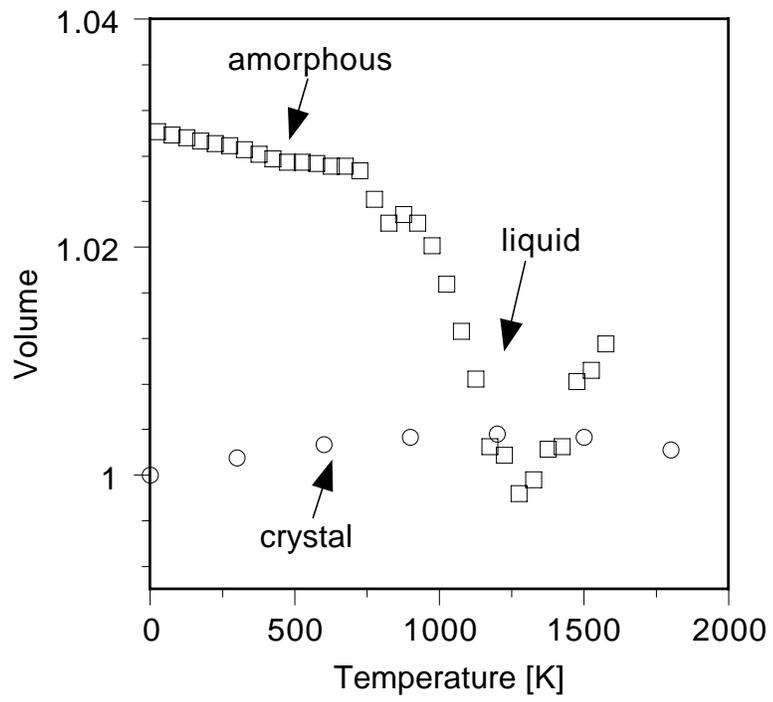

Figure 7.  Keblinski *at al.*

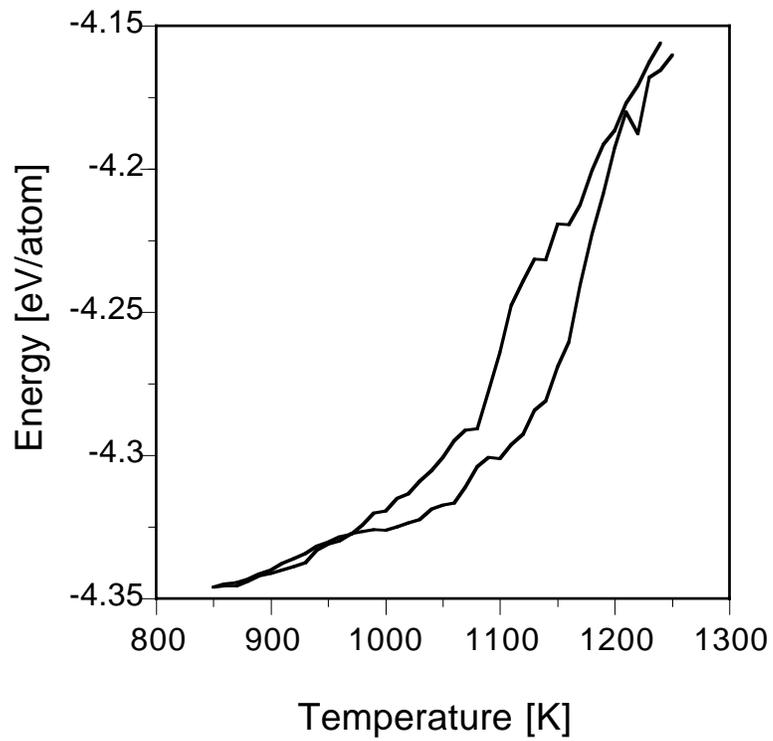

Figure 8.  Keblinski *at al.*

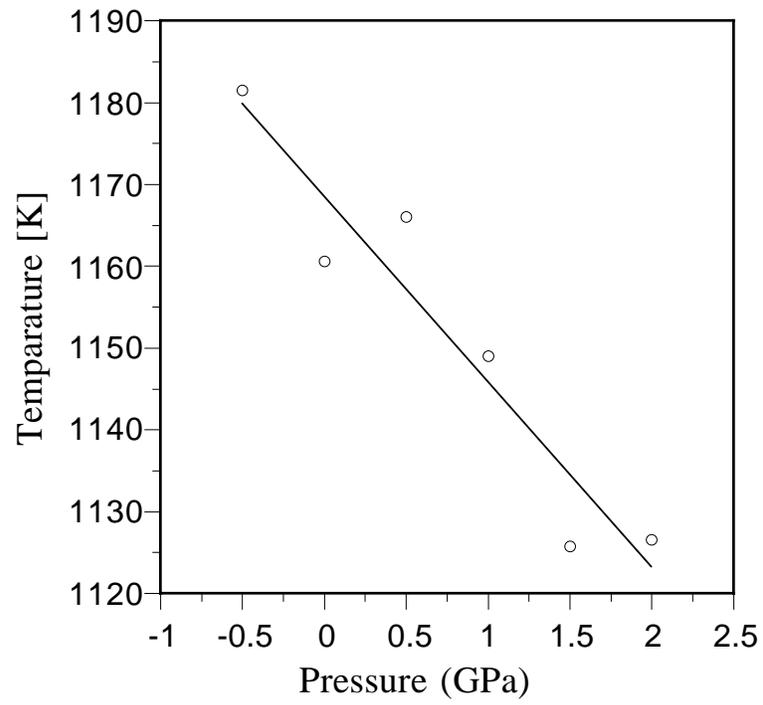

Figure 9.                               Keblinski *at al.*

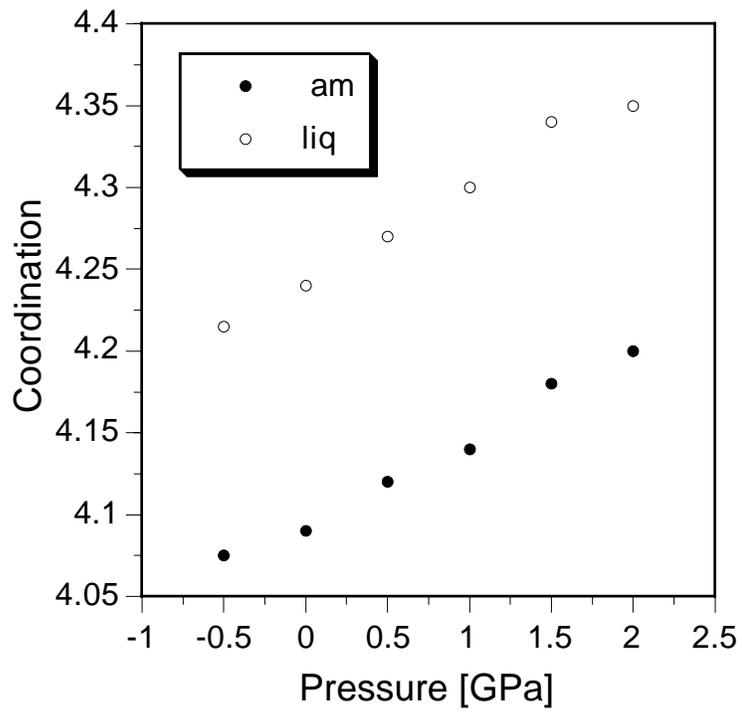

Figure 10.    Keblinski *at al.*

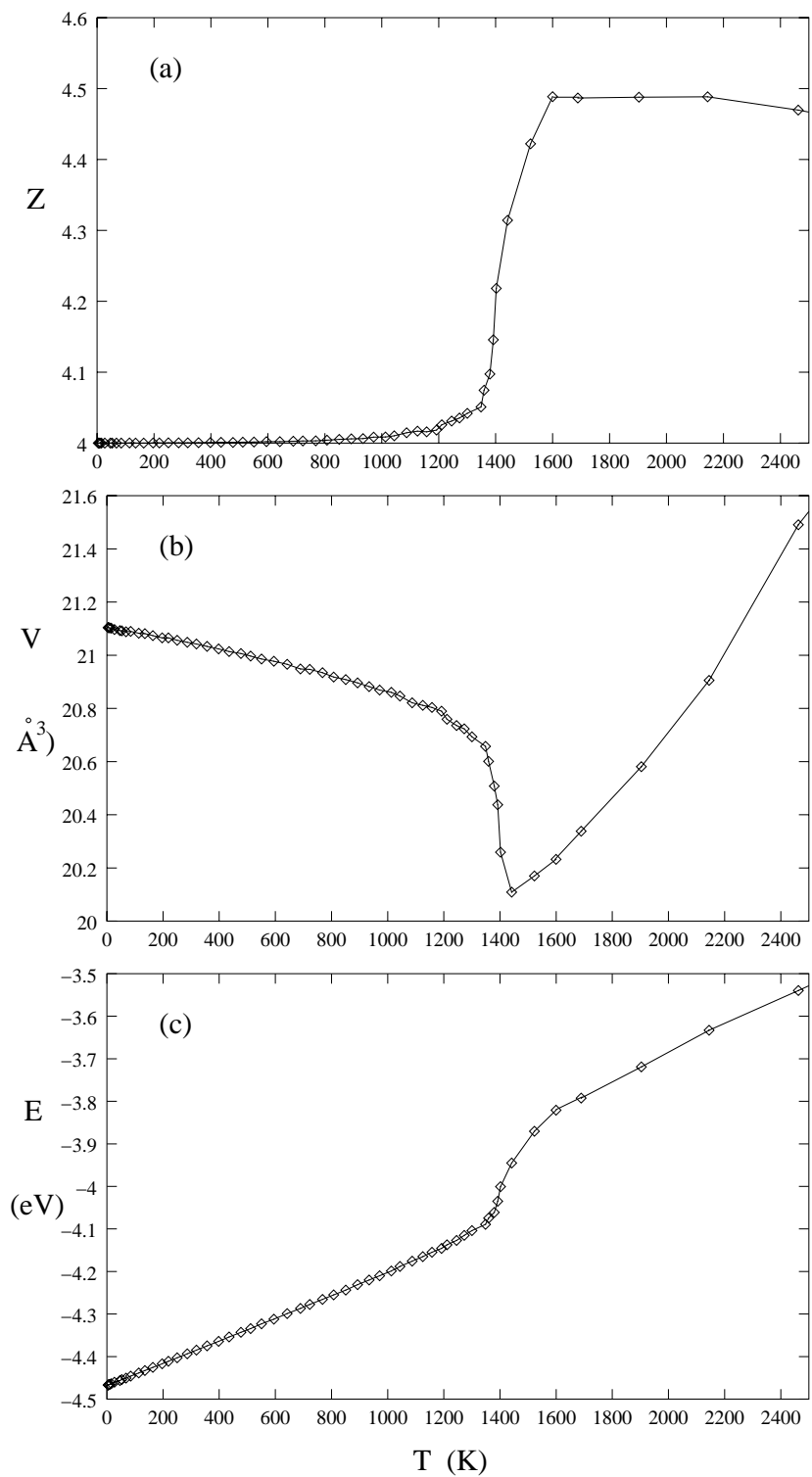

Figure 11.   Keblinski *at al*.

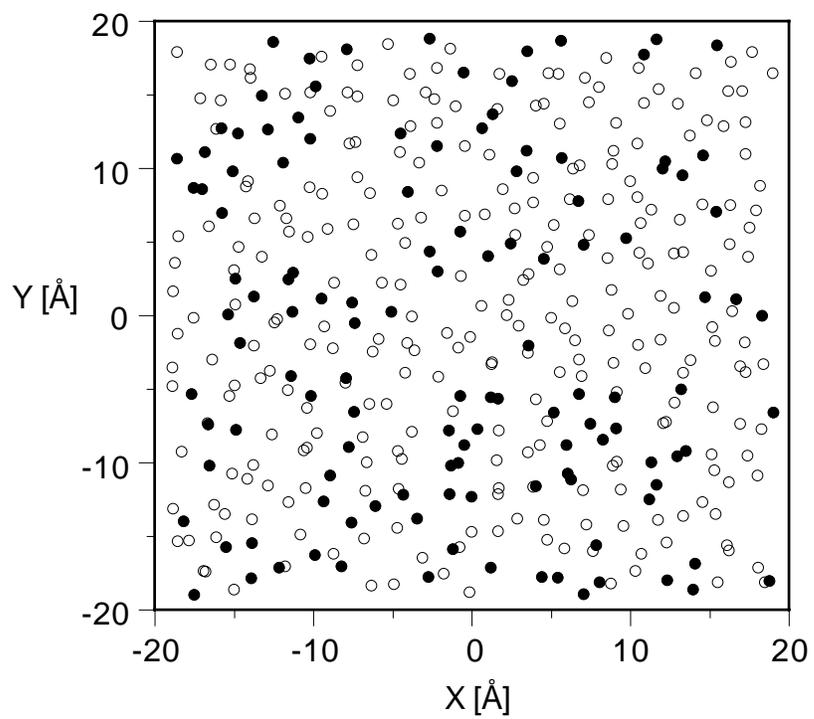

Figure 12.                                    Keblinski *at al.*

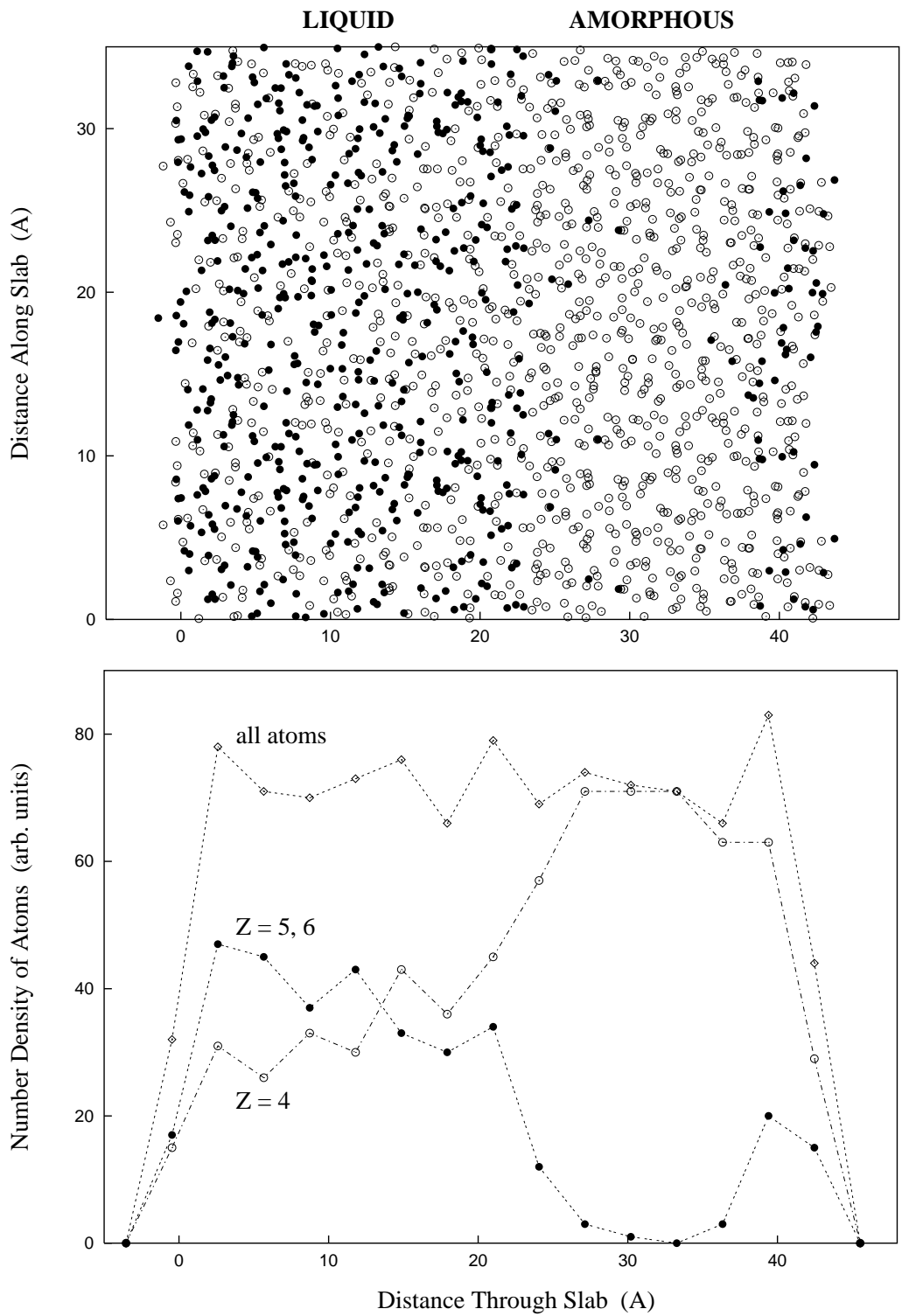

Figure 13. Keblinski *at al.*